\documentclass[aps,prl,preprint,superscriptaddress,showkeys]{revtex4-1}
\usepackage{graphicx}
\usepackage{color,soul}
\usepackage[sort&compress]{natbib}
\usepackage{dcolumn}% Align table columns on decimal point
\usepackage{bm}% bold math
\usepackage{sidecap}% 
\usepackage{amssymb,amsmath,subfigure}
\usepackage{verbatim}
\usepackage[export]{adjustbox}

\begin{document}

\title{2D vibrational properties of epitaxial silicene on Ag(111) \\probed by \textit{in situ} Raman Spectroscopy}

\author{Dmytro Solonenko}
\author{Ovidiu D. Gordan}
\affiliation{Technische Universit\"at Chemnitz, Reichenhainer Stra{\ss}e 70, 09126 Chemnitz, Germany}
\author{\\Guy Le Lay}
\affiliation{Aix-Marseille Universit\'e, CNRS, PIIM UMR 7345, 13397 Marseille Cedex, France}
\author{Dietrich R.T. Zahn}
\affiliation{Technische Universit\"at Chemnitz, Reichenhainer Stra{\ss}e 70, 09126 Chemnitz, Germany}
\author{Patrick Vogt}\email[]{patrick.vogt@physik.tu-berlin.de}
\affiliation{Technische Universit\"at Berlin, Hardenbergstra{\ss}e 36, 10623 Berlin, Germany}

%\author{}
%\thanks{}
%\affiliation{}


\date{\today}

\begin{abstract}
The two-dimensional silicon allotrope, silicene, could spur the development of new and original concepts in Si-based nanotechnology. Up to now silicene can only be epitaxially synthesized on a supporting substrate such as Ag(111). Even though the structural and electronic properties of these epitaxial silicene layers have been intensively studied, very little is known about its vibrational characteristics. Here, we present a detailed study of epitaxial silicene on Ag(111) using \textit{in situ} Raman spectroscopy, which is one of the most extensively employed experimental techniques to characterize 2D materials, such as graphene, transition metal dichalcogenides, and black phosphorous. The vibrational fingerprint of epitaxial silicene, in contrast to all previous interpretations, is characterized by three distinct phonon modes with $A$ and $E$ symmetries. The temperature dependent spectral evolution of these modes demonstrates unique thermal properties of epitaxial silicene and a significant electron-phonon coupling. These results unambiguously support the purely two-dimensional character of epitaxial silicene up to about 300$^{\circ}$C, whereupon a 2D-to-3D phase transition takes place.
\end{abstract}



\maketitle

%Introduction
The successful growth of silicene, the first purely synthetic elemental 2D honeycomb material, was reported in 2012 on a Ag(111) single crystal \cite{Vogt2012,Lin2012,Feng2012}, ZrB$_{2}$ \cite{Fleurence2012}, and later on Ir(111) \cite{Meng2013} templates. Yet, Ag(111) is by far the most used substrate material for the growth of silicene and it is the subject of most theoretical investigations. 
The archetype and best investigated epitaxial silicene structure on Ag(111), develops a $3\times3$ reconstruction coinciding with a $4\times4$ Ag(111) surface supercell (in short termed  $(3\times3)/(4\times4)$) \cite{Vogt2012}. Various experimental techniques have been applied to study this seminal silicene phase, such as scanning tunneling microscopy and spectroscopy (STM and STS) \cite{Vogt2012,Lin2012,Feng2012}, non-contact atomic force microscopy \cite{Resta2013}, as well as electron diffraction (LEED and RHEED), photoemission spectroscopy\cite{Vogt2012,Padova2013,Avila2013}, and large scale surface diffraction methods \cite{Takagi2015,Fukaya2013}. The suggested atomic model for this epitaxial silicene layer is supported by \textit{ab initio} calculations which also reveal a $sp^{2}$/$sp^{3}$ character of the single Si atoms \cite{Cahangirov2013,Vogt2012}. Apart from these achievements, a detailed picture of the vibrational properties of epitaxial silicene is still missing including a clear assignment of the phonon symmetries, their temperature dependence, and an experimental indication for the strength of the electron-phonon coupling. Raman spectroscopy allows these properties to be determined and has been one of the most extensively employed experimental techniques to study graphene \cite{Ferrari2007} and 2D layered materials like transition metal dichalcogenides (TMDCs) \cite{HongLi2012} and black phosphorous \cite{Xia2014}.  

Here, we report a thorough and comprehensive Raman study carried out \textit{in situ} to fully characterize $(3\times3)/(4\times4)$ epitaxial silicene on Ag(111) and to answer the crucial questions mentioned above. The presented results open important perspectives for future developments in the field of silicene research. The determined Raman fingerprint of epitaxial silicene on Ag(111), the related phonon symmetries and the temperature behaviour allow to compare these layers to silicene formed on other substrate materials and compare the silicene-substrate interaction. Furthermore, the results allow to identify silicene hidden underneath a protective capping layer, the properties of silicene multi-layer stacks and the analysis of structural changes upon functionalization of epitaxial silicene by atomic and molecular species.

%Experimental methods
Clean, well-ordered Ag(111) surfaces were prepared by Ar$^+$-bombardment (1.5 kV, $5\cdot10^{-5}$ mbar) and subsequent annealing at  $\sim530^\circ$C of (111)-oriented Ag single crystals under ultra-high vacuum conditions (base pressure $\sim2.0\cdot10^{-10}$ mbar). Si was deposited by evaporation from a directly heated Si-wafer piece, while the Ag sample was kept at a constant temperature, adjustable between 20$^\circ$C and $500^\circ$C with a precision of $\pm$ 10$^\circ$C. STM measurements were performed at room temperature in constant-current mode using an Omicron VT-STM with an electrochemically etched tungsten tip. The morphology of the Ag crystal and the differently prepared Si layers was probed by atomic force microscopy (AFM) (AFM 5420, Agilent Technologies Inc.) with commercial tips (curvature radius 10 nm) in non-contact mode, the symmetry was probed by low-energy electron diffraction (LEED). \textit{In situ} Raman measurements were performed in macro configuration, collecting the scattered light through the transparent port of a UHV chamber in front of the sample and recorded via a Dilor XY 800 monochromator. For excitation the Ar$^+$ 514.5 nm (2.4 eV) laser line was used with a power density of 10$^{3}$ W/cm$^{2}$ and an instrumental broadening of 2.5 cm$^{-1}$. Prior to the experiments we checked that heating effect of the silicene layer by the laser excitation can be neglected.

\begin{figure}
	\includegraphics[scale=1.5,center]{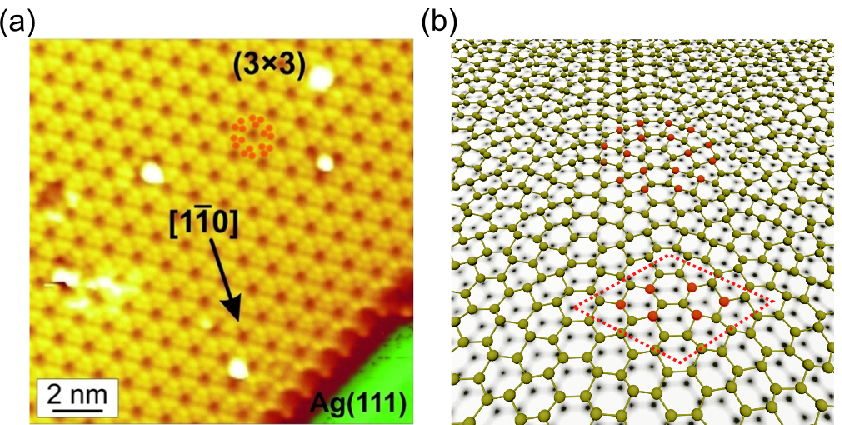}%
\caption{(a) STM image (U$_{bias}$ = -1.0 V, I = 1.08 nA) of epitaxial $(3\times3)/(4\times4)$ silicene (adapted from \citep{Vogt2014}). At the bottom right corner a part of the bare Ag(111) $1\times1$ surface can be seen. (b) Atomic ball-and-stick model for $(3\times3)/(4\times4)$ silicene (adapted from \citep{Vogt2012}). The $(3\times3)$ unit cell is indicated by a red rhombus.}\label{fig:Fig1}
\end{figure}

\begin{figure}
    \includegraphics[scale=0.5,left]{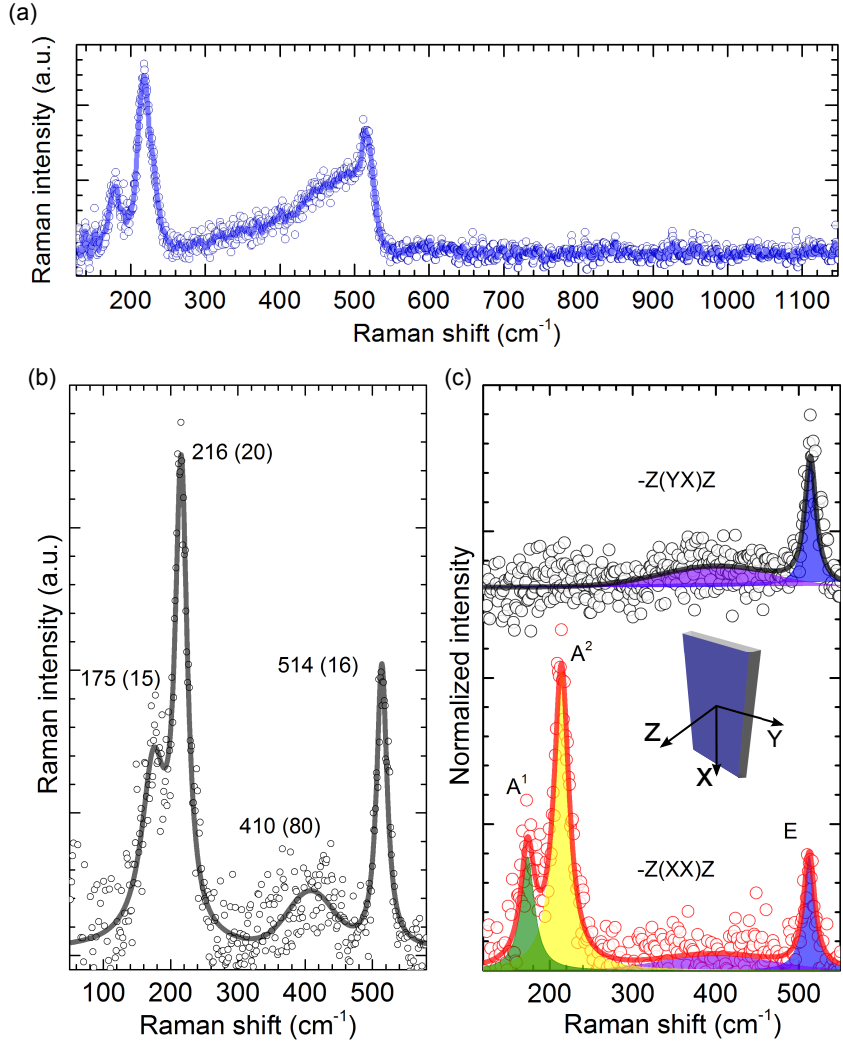}
\caption{(a) Raman spectrum of $(3\times3)/(4\times4)$ silicene up to 1200 cm$^{-1}$. The broad shoulder around 480 cm$^{-1}$ is associated with small amounts of co-existing a-Si. This component is subtracted in (b), clearly showing the phonon modes at 175 cm$^{-1}$, 216 cm$^{-1}$, 410 cm$^{-1}$, and 514 cm$^{-1}$ (FWHM in brackets). (c) Related Raman selection rules for parallel $\bar{z}(xx)z$ and crossed $\bar{z}(yx)z$ geometries.}\label{fig:Fig2}
\end{figure}

%Results and discussion
At a deposition temperature of approximately 220$^\circ$C the archetype $(3\times3)/(4\times4)$ 2D epitaxial silicene phase forms, showing in STM images (Fig.~\ref{fig:Fig1}(a), filled-states) the characteristic flower-like pattern of a regular and well-ordered $(3\times3)$ atomic structure, in agreement with our previous results \cite{Vogt2012}. Fig.~\ref{fig:Fig1}(b) shows the atomic ball-and-stick model for this layer. The Si atoms within the $(3\times3)$ unit cell show different displacements in the z-direction depending on their position. In the topographic STM image only the top atoms are imaged (red balls), producing the flower-like pattern. Within this structure most of the hexagons are oriented out-of-plane, while only the ones at the dark center of the flower pattern are in-plane.
Fig.~\ref{fig:Fig2}(a) shows an overview Raman spectrum of the epitaxial $(3\times3)/(4\times4)$ silicene phase recorded \textit{in situ}, at room temperature. The observed narrow Raman modes at 175 cm$^{-1}$, 216 cm$^{-1}$, and 514 cm$^{-1}$ underline the crystalline nature of epitaxial silicene. The presence of the broad bands at 350 cm$^{-1}$ and 480 cm$^{-1}$ is associated with the co-existence of smaller amounts of amorphous Si (a-Si) (see supplemental material) but not with relaxation of the momentum conservation law in the epitaxial silicene layer \cite{Cinquanta,Zhuang}.
The small intensity of the a-Si signature indicates that only a minor amount of a-Si is formed, probably at defective substrate areas. In order to obtain the pure spectral fingerprint of epitaxial $(3\times3)/(4\times4)$ silicene in Fig.~\ref{fig:Fig2}(b) the signature of a-Si was subtracted from the spectrum displayed in Fig.~\ref{fig:Fig2}(a).
In addition to the intense phonon modes at 175 cm$^{-1}$, 216 cm$^{-1}$, and 514 cm$^{-1}$, a weak broad band around 410 cm$^{-1}$ can be seen in the spectrum. This mode matches the spectral range, where a second-order phonon band of the intense mode at 216 cm$^{-1}$ can be expected.
In order to examine the nature of the observed modes at 175 cm$^{-1}$, 216 cm$^{-1}$, and 514 cm$^{-1}$, Raman selection rules were measured in two polarization geometries:  $\bar{z}(xx)z$ and $\bar{z}(yx)z$ (Porto notation), \textit{i.e.} parallel and crossed polarisations, respectively. The axes represent the sample coordinate system, where the z axis is normal to the substrate surface, while x (aligned with the Ag [-110] direction) and y represent the in-plane axes. In Fig.~\ref{fig:Fig2}(c) the Raman spectra for both geometries are shown together with the fitted line shapes of the single components. The two Raman modes at 175 cm$^{-1}$ and 216 cm$^{-1}$ are clearly visible in parallel geometry but they disappear completely in crossed geometry while the mode at 514 cm$^{-1}$, as well as the broad band at 410 cm$^{-1}$ are present in both polarization geometries. Such behavior is governed by different symmetries of the lattice vibrations associated with these modes. The $(3\times3)/(4\times4)$ structure of the epitaxial silicene layer belongs to the C$_{3}$ symmetry point group \cite{Pflugradt2014}, which possesses three possible phonon symmetries: $A(z)$, $E(x)$, and $E(y)$.
\[ A(z) = \begin{pmatrix}
a & 0 & 0 \\
0 & a & 0 \\
0 & 0 & b
\end{pmatrix}
;~
E(x) = \begin{pmatrix}
c & d & e \\
d & -c & f \\
e & f & 0 
\end{pmatrix};
\]
\[E(y) = \begin{pmatrix}
d & -c & -f \\
-c & -d & e \\
-f & e & 0
\end{pmatrix}
\]
\\
It is found that $A$ modes appear only in the parallel geometry while the $E$ modes appear in both geometries. Thus, the polarization-dependent Raman results conclusively show that the modes at 175 cm$^{-1}$ and 216 cm$^{-1}$ are fully symmetric vibrations and are assigned to an $A$ symmetry, denoted as $A^{1}$ and $A^{2}$, respectively. A mode similar to $A^{2}$ was also observed by \textit{in situ} Raman measurement at 77 K and related to light scattering from domain boundaries \cite{Zhuang}. Hence it was assigned to a `$D$ band', by analogy to graphene. However, the very clear assignment of $A^{1}$ and $A^{2}$ to an $A$ symmetry rules out such explanations, since in this case a polarization dependence should not be observed, because domain boundaries break the long-range translational symmetry in the 2D crystal. The Raman band at 410 cm$^{-1}$ shows a weaker polarization dependence expected for a second order phonon. The mode at 514 cm$^{-1}$ is assigned to an $E$ symmetry based on the observed polarization dependence. This phonon mode was also reported in previous \textit{ex situ} and \textit{in situ} Raman measurements at 516 cm$^{-1}$ and 530 cm$^{-1}$, respectively \cite{Zhuang,Cinquanta}. The blueshift in the latter case can be partly related to the lower temperature of 77 K during the measurements but the total shift is too large to relate it solely to thermal effects. In any case, our results provide now experimental evidence for the $E$ symmetry of this mode.

It is remarkable that the Raman signature of epitaxial $(3\times3)/(4\times4)$ silicene is dominated by phonons with an $A$ symmetry. $A$ modes do not exist in diamond-like Si \cite{Richter1981,Menendez1984} or other bulk Si allotropes. Due to the 2D nature of the silicene lattice, the translational symmetry is broken in one direction (perpendicular to the lattice plane). This lifts the phonon triple degeneracy at the $\Gamma$ point, as it was shown by theoretical calculations for the phonon dispersion of free-standing silicene  \cite{Yan2013,Gori2014,Li2013}, where the former TO phonon mode of bulk Si shifts to lower energy. Moreover, according to group theory analysis \cite{Ribeiro2015}, free-standing silicene has Raman-active modes of $A$ and $E$ symmetries, unlike bulk Si, which has only one zone-centre Raman mode of $F$ symmetry. We stress that the appearance of $A$ modes in the case of $(3\times3)/(4\times4)$ silicene on Ag(111) reflects the 2D character of epitaxial silicene. These $A$ modes are associated with an out-of-plane displacement of Si-atoms, a vibrational mode that is supported by the buckling and the out-of-plane oriented hexagons of the $(3\times3)/(4\times4)$ atomic structure (Fig.~\ref{fig:Fig1}(b)). Such modes are not observed for the fully-planar structure of graphene. 

The symmetries of the observed Raman modes are in good agreement with theoretical expectations for free-standing silicene, whilst their frequencies show much less agreement with available calculations of the phonon dispersion \cite{Yan2013,Gori2014,Li2013}. DFT calculations suggest the presence of three optical phonon branches at $\Gamma$: a ZO mode, related to out-of-plane optical phonons, and energetically degenerate TO and LO phonons. While the position of the ZO branch at $\Gamma$, is close to the experimentally observed $A$ modes, the degenerate TO/LO branches (562 cm$^{-1}$ \cite{Yan2013}, $\sim$ 550 cm$^{-1}$ \cite{Gori2014}, $\sim$ 556 cm$^{-1}$ \cite{Li2013}) do not match the $E$ mode of epitaxial silicene (514 cm$^{-1}$). Such a discrepancy can be explained by the significant interaction between the Si adlayer and the substrate, the different atomic arrangement caused by the superstructure formation of $(3\times3)/(4\times4)$ silicene, and the related modified hybridization state with respect to free-standing silicene \cite{Cahangirov2013}. 

We emphasize that the Raman observations give no indication for Si-Ag vibrational modes, resulting from inter-atomic bonding between these two species. Theoretical calculations show that a Ag-Si related mode should evolve at about 90 cm$^{-1}$ \cite{Chou1988}. While it was suggested that Si-Ag bonds, or even alloying, could play a role for some of the Si phases on Ag(111) \cite{Prevot2014}, this is not at all observed for $(3\times3)/(4\times4)$ silicene/Ag(111).

\begin{figure}[h]
   \includegraphics[scale=0.8, center]{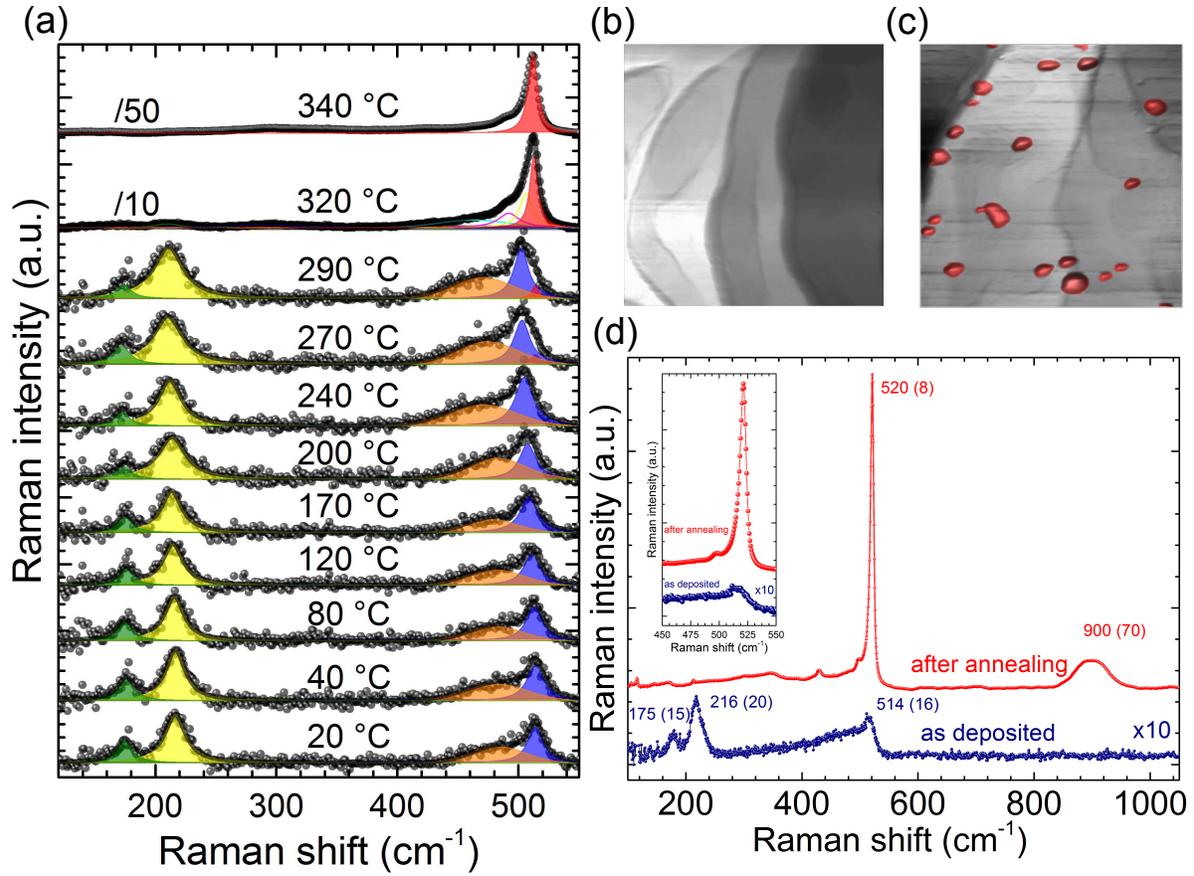}%
\caption{(a) Series of Raman spectra of $(3\times3)/(4\times4)$ silicene measured at increasing temperatures. %Around 300~$^\circ$C the spectral line shape changes dramatically as a result of a phase transition due to a dewetting process. 
(b) AFM image (2 $\mu$m $\times$ 2 $\mu$m) of the initial Ag(111) surface. (c) AFM image (2 $\mu$m $\times$ 2 $\mu$m) after annealing the $(3\times3)/(4\times4)$ silicene layer to 500$^\circ$C with small islands (in red). (d) Overview Raman spectra of the epitaxial silicene before and after annealing to 500$^\circ$C both measured at room temperature (FWHM in brackets). Inset: detailed Raman spectra (450 - 550 cm$^{-1}$).}\label{fig:Fig3}
\end{figure}

Fig.~\ref{fig:Fig3}(a) shows a series of Raman spectra measured during the annealing of the epitaxial $(3\times3)/(4\times4)$ silicene layer on Ag(111). The Raman spectra were recorded in incremental temperature steps from room temperature to 500$^\circ$C. For temperatures up to $\sim300^\circ$C the overall spectral line shape remains unchanged and all modes shift almost equally towards lower wavenumbers with increasing temperature. Such temperature dependence is qualitatively very similar to the ones of graphene \cite{Calizo2007}, MoS$_{2}$ \cite{Sahoo2013}, or bulk Si \cite{Menendez1984,Hart1970} where the Raman modes, related to optical phonons, shift  towards lower energies. 
At a temperature slightly above 300$^\circ$C the spectrum undergoes a peculiar change: the $E$ mode, which is found at 514 cm$^{-1}$ at room temperature, broadens considerably as a result of a new component appearing at around 520 cm$^{-1}$ (measured at room temperature). For higher temperatures this new mode increases and finally dominates the spectra. These results demonstrate that a structural phase transition takes place at temperatures around 300$^\circ$C. According to the Raman spectra, the silicene layer transforms into a phase which exhibits a bulk Si-like Raman spectrum characterized by the L(T)O mode at 520 cm$^{-1}$ with an asymmetric shoulder on the lower energy side. AFM images acquired at room temperature after the annealing procedure to 500$^\circ$C (Fig.~\ref{fig:Fig3}(c)) show the formation of small islands with an average size of (10 $\pm$ 8) nm, distributed over the Ag surface. These islands are not observed on the initial Ag(111) surface (Fig.~\ref{fig:Fig3}(b)) or after formation of the epitaxial silicene layer and are, thus, assigned to Si crystallites or Si nanoparticles. This assignment is justified since Si nanoparticles show a L(T)O mode around 520 cm$^{-1}$ with an asymmetric shoulder, caused by spatial phonon confinement in particles smaller than 7 nm \cite{Campbell1986}, while their size distribution explains the broad linewidth of this mode. In Fig.~\ref{fig:Fig3}(d) the room temperature Raman spectrum of the same sample is shown in a spectral range between 100 cm$^{-1}$ and 1000 cm$^{-1}$. Note that in this instance, \textit{i.e.} after formation of Si crystallites, a broad band (2nd-order of the L(T)O mode) centered around 950 cm$^{-1}$ can be seen, which is not the case for the epitaxial silicene layer. This means that the occurrence of this 2nd-order band is indicative for the formation of Si crystallites.

\begin{figure}
    \includegraphics[scale=0.8,center]{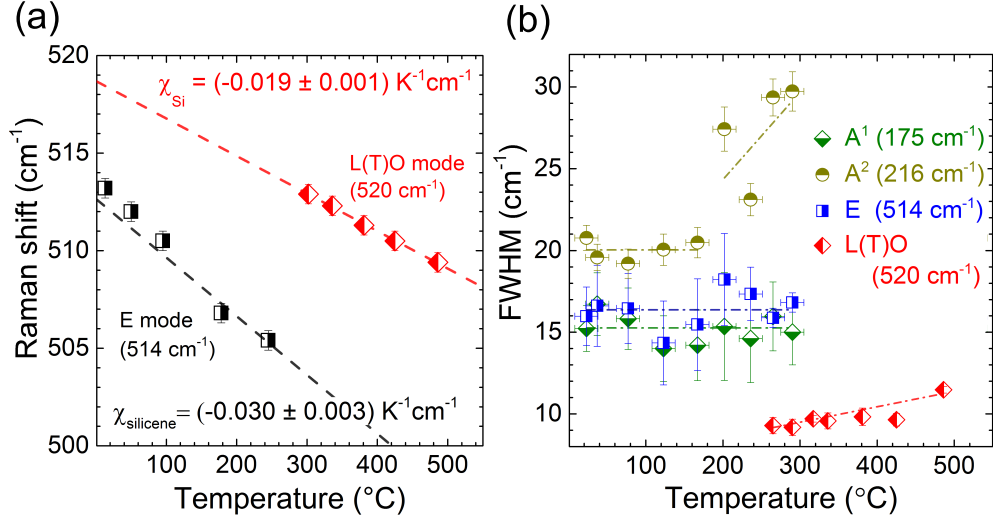}%
\caption{(a) Raman shift of the E mode of $(3\times3)/(4\times4)$ silicene  as a function of temperature. (b) Full width at half maximum (FWHM) of the silicene $A^{1}$, $A^{2}$, and $E$ and the Si L(T)O Raman modes as a function of temperature (with error bars). Experimental points are linearly fitted (dash-dot lines).}\label{fig:Fig4}
\end{figure}

	Additional evidence for a phase transition can be found by following the behavior of the Raman modes upon heating. The temperature-dependent position of the $E$ mode is depicted in Fig.~\ref{fig:Fig4}(a). The shift with increasing temperature allows the determination of its thermal coefficient, having a value of $\chi_{silicene} = (-0.030 \pm  0.003)$ K$^{-1}$cm$^{-1}$. Instead, the thermal coefficient of the L(T)O phonon of the Si nanocrystallites after the phase transition is found to be $\chi_{Si} = (-0.019 \pm 0.003)$ K$^{-1}$cm$^{-1}$, in good agreement with values reported for diamond-like bulk Si \cite{Menendez1984,Hart1970,Perichon1999}. Note, the determination of the thermal coefficient of silicene is not dependent on absolute temperature and hardly influenced by the external factors such as the underlying Ag substrate. The shift of the other Raman modes of epitaxial silicene (not shown) is similar to the one of the $E$ mode. This reveals that the thermal coefficient of the Raman modes of epitaxial silicene on Ag(111) markedly differs from the one of bulk-like Si. A similar difference is found between graphene (2D) and diamond (3D), where the zone-centre $G$ mode of graphene has a higher thermal coefficient compared to the diamond phonon mode \cite{Calizo2007}. 
	
	Further information is gained upon studying the evolution of the FWHM of the silicene Raman modes (Fig.~\ref{fig:Fig4}(b)). While the widths of the $A^{1}$ ($\sim$15 cm$^{-1}$) and the $E$ ($\sim$16 cm$^{-1}$) modes remain unchanged up to 300$^\circ$C, the $A^{2}$ mode shows a constant width ($\sim$20 cm$^{-1}$) for temperatures up to 220$^\circ$C. For temperatures exceeding 240$^\circ$C the $A^{2}$ mode starts to broaden simultaneously with the appearance of the L(T)O mode (originating from the Si nanoparticles). Moderate dewetting of Si atoms from the Ag surface creates local disorder \cite{Acun2013}, which could be related to the $A^{2}$ mode broadening. Remarkably, only the $A^{2}$ mode responds to this structural change, showing similarities to the broadening of the $D$ band of graphene upon temperature increase \cite{Ferrari2007}. At temperatures above 300$^\circ$C, we can only follow the FWHM of the Raman mode at 520 cm$^{-1}$, as the Raman modes of silicene have disappeared. Its linewidth (FWHM at room temperature: $\sim$6 cm$^{-1}$) evolves in a manner, expected for nanocrystalline Si \cite{Mishra2000}.
	
	 Overall, all observed Raman modes for epitaxial $(3\times3)/(4\times4)$ silicene show a markedly larger broadening than, for example, the L(T)O mode of bulk Si ($\sim$2.4 cm$^{-1}$). This broadening could be caused by different effects such as a loss of crystallinity and lattice disorder, confinement effects due to small domain sizes or a considerably strong electron-phonon coupling (EPC). However, the very clear polarization dependence of the phonon modes (described above) and the well defined layer periodicity (according to LEED and STM) demonstrate that disorder as an explanation of the phonon broadening can be neglected. Phonon confinement could occur if the average domain size of the epitaxial silicene is in the range of a few nm ($\sim 7$ nm for Si allotropes \cite{Mishra2001}). This would lead to a lifting of momentum conservation which disagrees again with the clear fulfilment of the Raman selection rules. Additionally, the lifting of momentum conservation implies an averaging over the phonon dispersion curves from the $\Gamma$ point towards the Brillouin zone edge. The branches of optical phonons usually have a negative slope around the $\Gamma$ point and such averaging would cause an asymmetry of the phonon modes in the case of a significant slope. Since a peak asymmetry is not observed (as it would be expected for Si allotropes \cite{Gori2014}), phonon confinement cannot explain the phonon broadening of epitaxial silicene either. Ultimately, the large linewidth of the silicene phonon modes hints towards a strong electron-phonon coupling. A very similar effect was shown to introduce a significant broadening of $G$ and $G$' phonon bands of graphene and carbon nanotubes up to 11-13 cm$^{-1}$ \cite{Ferrari2007,Lazzeri2006}. By analogy with the constant width of the $G$ mode of graphene versus temperature, we also observe a constant linewidth with temperature of the $A$ and $E$ modes, which supports the existence of strong coupling. Significant electron-phonon coupling at the $\Gamma$ point was also predicted theoretically for free-standing silicene \cite{Li2013}.
	
To summarize, we have revealed and identified the intrinsic Raman spectral signatures of epitaxial silicene on Ag(111). The properties, such as the phonon symmetries and their thermal coefficients, confirm the 2D nature of this first synthetic group IV elemental 2D material beyond graphene. However, silicene is not just a Si-based graphene copy, since its intrinsic buckling, modified by the silver (111) substrate, alters the vibrational structure in contrast to the flat case. Furthermore, \textit{in situ} Raman spectroscopy proves that the formation and the structural stability of the epitaxial silicene monolayer are limited within a restricted temperature range. At 300$^\circ$C a dewetting transition takes place and diamond-like Si nanocrystals are formed.

\begin{acknowledgments}
This work was financially supported by the Deutsche Forschungsgemeinschaft (DFG) under Grant No. VO1261/3-1 and VO1261/4-1, the International Research Training Group (GRK 1215), jointly sponsored by the DFG and the Chinese Ministry of Education, and the 2D-NANOLATTICES project within the 7th Framework Programme for Research of the European Commission, under FET-Open Grant No. 270749.
\end{acknowledgments}

\bibliographystyle{apsrev}

\newpage

\end{document}